%% file: paper.tex
\documentclass[a4paper,11pt]{article}
\pdfoutput=1 
\usepackage{aas_macros,braket}
\usepackage{jcappub,natbib} 
\usepackage[T1]{fontenc} 

\title{Primordial trispectra and CMB spectral distortions}

\author[a,b]{Nicola Bartolo,}
\author[a,b]{Michele Liguori,}
\author[c]{and Maresuke Shiraishi}

\affiliation[a]{Dipartimento di Fisica e Astronomia ``G. Galilei'', Universit\`a degli Studi di Padova, via Marzolo 8, I-35131, Padova, Italy}
\affiliation[b]{INFN, Sezione di Padova, via Marzolo 8, I-35131, Padova, Italy}
\affiliation[c]{Kavli Institute for the Physics and Mathematics of the Universe (Kavli IPMU, WPI), UTIAS, The University of Tokyo, Chiba, 277-8583, Japan}

\emailAdd{nicola.bartolo@pd.infn.it}
\emailAdd{michele.liguori@pd.infn.it}
\emailAdd{maresuke.shiraishi@ipmu.jp}

\abstract{
We study the $TT\mu$ bispectrum, generated by correlations between Cosmic Microwave Background temperature (T) anisotropies and chemical potential ($\mu$) distortions, and we analyze its dependence on primordial local trispectrum parameters $g_{\rm NL}$ and $\tau_{\rm NL}$. We cross-check our results by comparing the full bispectrum calculation with the expectations from a general physical argument, based on predicting the shape of $\mu$-T correlations from the couplings between short and long perturbation modes induced by primordial non-Gaussianity. We show that {\em both} $g_{\rm NL}$ and $\tau_{\rm NL}$-parts of the primordial trispectrum source a non-vanishing $TT\mu$ signal, contrary to the $\mu\mu$ auto-correlation function, which is sensitive only to the $\tau_{\rm NL}$-component. A simple Fisher matrix-based forecast shows that a futuristic, cosmic-variance dominated experiment could in principle detect $g_{\rm NL} \sim 0.4$ and $\tau_{\rm NL} \sim 40$ using $TT\mu$.
}

\begin{document}

\input{macros.tex}

\begin{flushright}
{\small IPMU15-0192}
\end{flushright}

\maketitle
\flushbottom

\section{Introduction}

Measurements of primordial non-Gaussianity (NG) are a powerful way to 
understand the physical processes which gave origin to primordial cosmological 
perturbations. They provide information about such processes which is 
complementary to what can be extracted from power spectrum 
alone. If 
 we focus on inflationary scenarios, all relevant NG information is 
generally contained in the bispectrum (three-point function in Fourier space)
and trispectrum (four-point function in Fourier space) of the primordial 
 fluctuation field.
Both the functional form (``shape'') and strength of these 
signals are model dependent, therefore constraints on different inflationary 
scenarios can be obtained by fitting their predicted bispectrum 
and trispectrum shapes to the data, and extracting the corresponding 
amplitude parameters $\fnl$ (for the bispectrum), $\gnl$ and $\taunl$ 
(for the trispectrum).

The first inflation-motivated primordial NG model to be considered in 
the literature \cite{Salopek:1990jq, Gangui:1993tt} was the so called ``local model'', which is characterized 
by the following ansatz in real space:
\begin{eqnarray}
  \zeta(\mathbf{x}) = \zeta^{\rm G}(\mathbf{x}) + \frac{3}{5} \fnl \left( \zeta^{\rm G}(\mathbf{x}) - \left\langle \zeta^{\rm G}(\mathbf{x}) \right \rangle \right)^2 
  + \frac{9}{25} \gnl \left( \zeta^{\rm G}(\mathbf{x}) \right)^3 \; ,
\end{eqnarray}
where $\zeta$ is the primordial curvature perturbation field,
$\zeta^{\rm G}$
is its Gaussian (G) part and the NG components are local 
functionals of the G part. One can also consider models in which $\zeta^{\rm G}$ is modulated 
by a second, uncorrelated, Gaussian field $\sigma$, giving rise to a ``$\taunl$ trispectrum'' \cite{Smith:2015uia}:
\begin{equation}\label{eqn:taunl}
\zeta(\bx) = \zeta^{\rm G}(\bx) + \sqrt{\taunl} \sigma(\bx)\zeta^{\rm G}(\bx) \; .
\end{equation}
As we just mentioned, different primordial models 
can generate a large variety of different bispectrum and trispectrum 
shapes, and to each of them correspond different NG amplitudes. 
The focus of this paper will however be specifically on local-type 
bispectra and trispectra, which are produced by a primordial curvature perturbation field
expressed in the form above.~\footnote{Therefore, since there is no room for confusion, 
we will simply refer to our NG parameters as $\fnl$ and $\gnl$, omitting the label ``local''.}

Currently, the most stringent constraints on primordial NG come from $\Planck$  
measurements of the Cosmic Microwave Background (CMB) temperature and polarization bispectra 
and trispectra \cite{Ade:2013ydc, Feng:2015pva, Ade:2015ava}. 
For the local shape, they are $\fnl = 2.5 \pm 5.7$ (68 \% CL),
$\gnl = (-9.0 \pm 7.7)\times 10^{4}$ (68 \% CL) \cite{Ade:2015ava}, $\taunl < 2800$ (95 \% CL) \cite{Ade:2013ydc}. 
While being very tight, and representing, as emphasized in $\Planck$ papers, the 
highest precision test to date of the standard single-field slow-roll paradigm, these results by no mean rule out 
more complex multi-field models. In absence of a clear detection, if we want to convincingly discriminate between 
single and multi-field scenarios, it would in fact be necessary to find a test producing at least one order of magnitude, 
if not better, improvements in $\fnl$ error bars. This would probe a range of amplitudes which 
is at the level of the standard single-field slow-roll prediction, $\fnl \sim \epsilon$, where $\epsilon \sim 10^{-2}$ is the 
slow-roll parameter. Moreover, popular multi-field scenarios, such as the curvaton model, naturally predict a lower bound 
of order unity for $|\fnl|$, so that constraining this parameter to take much smaller values would effectively rule them out. Significant improvements in trispectrum constraints would also provide crucial information, allowing to further discriminate between competing scenarios for the origin of cosmic structures. 
In particular, sizable squeezed trispectra $\gnl$ and $\taunl$ can arise only within multi-field models of inflation. If there is a nonvanishing local bispectrum then there must be a trispectrum with $\taunl \geq (6 f_{\rm NL}/5)^2$, according to the Suyama-Yamaguchi relation~\cite{Suyama:2007bg}. Moreover there are inflationary scenarios where the trispectrum has larger signal-to-noise than the bispectrum. For example, this is the case of some curvaton~\cite{Sasaki:2006kq,Byrnes:2006vq} or other multi-field models~\cite{Byrnes:2010em,Ichikawa:2008ne}, where, for some parameter values, significant $\gnl$ and $\taunl$, respectively, can be generated along with a small $f_{\rm NL}$. Example of technically natural models where the trispectrum has larger signal-to-noise~\cite{Senatore:2010jy,Baumann:2011nk,Bartolo:2010di} do exist. This can happen also in multi-field models where the observed curvature perturbation is modulated by an uncorrelated field, such as those parametrized by Eq.~(\ref{eqn:taunl}). These models are characterized by a vanishing bispectrum, thus leaving the $\taunl$-trispectrum as the main NG signature. 
$\Planck$ has nearly saturated the maximum amount of information on NG parameters that can be extracted from CMB temperature and polarization. Even an ideal, cosmic variance dominated experiment, could 
not improve on current NG constraints by more than a factor $\sim 2-3$. It is then clear that, while CMB anisotropies 
have been the main driver for experimental NG studies up to now, we will have to turn to different observables in the future, 
if we hope to achieve the desired order of magnitude(s) leap. 

A natural attempt, in this respect, is to look at Large Scale Structure statistics and forthcoming Euclid data \cite{Laureijs:2011gra}. 
Expectations of future improvements in this case rely at the moment mostly on measurements of scale-dependent halo bias 
in the 2-point function \cite{2012MNRAS.422.2854G}, but, even in the most optimistic picture, forecasted error bars are far 
from allowing to explore the $\fnl \ll 1$ regime we are ultimately interested in, as well as from producing 
order-of-magnitude improvements in trispectrum parameters.

If we look at more futuristic scenarios, three main approaches have been proposed to achieve the desired sensitivity 
on NG parameters. One is to look at the higher-order correlation functions of full-sky 21-cm radiation surveys, in the 
redshift range $30 < z <100$ (e.g., \cite{Cooray:1999kg,Cooray:2004kt,Cooray:2008eb, Pillepich:2006fj,Munoz:2015eqa, Shimabukuro:2015iqa}). Another possibility is to study scale-dependent bias 
in future radio surveys probing high redshifts (e.g., \cite{Giannantonio:2011ya, Maartens:2012rh, Byun:2014cea, Raccanelli:2015oma}). The third approach, which we focus on in this paper, was recently introduced 
in \cite{Pajer:2012vz}. It consists in measuring cross-correlations between CMB chemical potential ($\mu$) spectral distortions, arising from 
dissipation of acoustic waves in the primordial photon-baryon plasma, and temperature anisotropies. As originally pointed out 
by the authors of \cite{Pajer:2012vz}, the $\mu T$ correlation probes the local bispectrum at wavenumbers 
$50 \, {\rm Mpc}^{-1} \lesssim k \lesssim 10^4 \, {\rm Mpc}^{-1}$, i.e. on scales which are unaccessible by CMB temperature or 
polarization anisotropies, or by any other cosmological probe, including future galaxy and 21-cm surveys. An ideal,  
cosmic-variance dominated experiment could extract a very large number of modes in this range of scales, allowing in 
principle constraints on $\fnl \lesssim 10^{-3}$. Moreover, it was also shown that, by the same reasoning, cosmic variance 
dominated $\mu \mu$ measurements could constrain $\taunl$ with an exquisite level of precision as well. These original 
findings have been followed by further studies from several groups, showing that $\mu T$ correlations could 
be used to study several other NG signatures besides standard local-type NG \cite{Pajer:2012vz,Ganc:2012ae,Biagetti:2013sr, Miyamoto:2013oua, Kunze:2013uja, Ganc:2014wia, Ota:2014iva, Emami:2015xqa, Shiraishi:2015lma}. Recent $\fnl$ constraints with this technique were obtained in \cite{Khatri:2015tla} using $\Planck$ 
data.

One interesting primordial NG parameter, that $\mu T$ and $\mu \mu$ correlations are unable to determine, is the $\gnl$ trispectrum 
amplitude. It can in fact be shown (see also Sec.~\ref{sec:fisher}) that $\mu \mu$ correlations are not sensitive to $\gnl$-type local 
NG. In this paper, we will point out that $\gnl$ can however still be measured by going beyond two-point correlations and 
using the $TT\mu$ {\em bispectrum}. We will then show that $TT\mu$ allows to measure not only $\gnl$, but also to extract 
additional information on $\taunl$. By a simple Fisher matrix forecast, we will finally conclude that $TT\mu$ bispectrum estimates 
could in principle allow a sensitivity $\Delta \gnl = {\cal O}(0.1)$ in the ideal, cosmic variance dominated case. Such exquisite precision can be 
attained, as usual in this approach, thanks to the very large number of primordial bispectrum modes that are contained in the $TT\mu$ 
  three-point function.

The plan of this paper is as follows: in Sec.~\ref{sec:phys} we start with a simplified calculation, aimed at putting in evidence the 
physical mechanism which produces the $\gnl$ and $\taunl$ dependencies in the $TT\mu$ bispectrum. We then perform the 
full calculation in Sec.~\ref{sec:full}, finding a nice agreement with the previous result, and show some $\gnl$ and $\taunl$ 
Fisher-based forecasts in Sec.~\ref{sec:fisher}, before reporting our conclusion in Sec.~\ref{sec:conclusions}.

\section{Preliminary calculation}\label{sec:phys}

Here we show a preliminary calculation of the $TT \mu$ signal, using a configuration space approach originally introduced in \cite{Emami:2015xqa}, 
where it is explained in detail. 
The idea is to estimate the expected correlations between $\mu$ and T via a short-long mode splitting of the primordial fluctuation field.
We are in fact interested here in CMB distortions arising from dissipation of primordial perturbations on small scales. These will be  
proportional to the primordial {\em small scale} power. For Gaussian initial conditions, different small scale patches are uncorrelated, and 
the average distortion will be the same everywhere. If, however, we are in presence of NG initial conditions correlating large and small scales, 
such as local-type NG, the average small-scale power will vary from patch to patch, and it will be correlated with curvature fluctuations on large 
scales. We can thus infer the expected fluctuations in the $\mu$  (and $y$) distortions parameter by evaluating the contributions to 
small scale power, coming from correlations with long wavelength modes. In this framework, let us consider a NG primordial perturbation field, 
with non-zero $\gnl$, while keeping $\fnl=0$ and $\taunl=0$:
\begin{equation}
\label{zetacubic}
\zeta(\mathbf{x}) = \zeta^{\rm G}(\mathbf{x}) + \frac{9}{25} \gnl \left( \zeta^{\rm G}(\mathbf{x})\right)^3 \; .
 \end{equation}
 Let us split the curvature perturbation $\zeta^{\rm G}({\bf x})$ into short and long wavelength parts,  
$\zeta^{\rm G}({\bf x})=\zeta^{\rm G}_S({\bf x})+\zeta^{\rm G}_L({\bf x})$, and similarly for $\zeta({\bf x})$. Using this split into Eq.~(\ref{zetacubic}) we can read the corresponding short and long wavelength contributions to $\zeta({\bf x})$. The dominant terms are 
\begin{eqnarray}
  \zeta({\bf x}) &=& \zeta_S({\bf x})+\zeta_L({\bf x}) \nonumber \\ 
  &=&  \zeta^{\rm G}_S({\bf x})+\zeta^{\rm G}_L({\bf x})+\frac{27}{25} g_{\rm NL} \zeta^{\rm G}_S({\bf x}) \left(\zeta^{\rm G}_L({\bf x}) \right)^2\, , 
\end{eqnarray}
so that the small-scale curvature perturbation modulated by the long-wavelength modes is given by 
\begin{equation}
\zeta_S({\bf x})=\zeta^{\rm G}_S({\bf x})\left[1+\frac{27}{25} g_{\rm NL} \left (\zeta^{\rm G}_L({\bf x}) \right)^2\right]\, . 
\end{equation}
The fractional change in small-scale power due to the long-wavelength mode is therefore 
\begin{equation}
\frac{\delta \langle \zeta^2 \rangle }{\langle \zeta^2\rangle }\simeq \frac{\delta \mu}{\mu}\simeq \frac{54}{25} g_{\rm NL} \left( \zeta^{\rm G}_L({\bf x}) \right)^2\, .
\end{equation}
As written in the second equality the fractional change in small-scale power determines the fractional change in the $\mu$ type distortions, since the average 
$\mu$ distortions are given by 
\begin{equation}
\label{muaverage}
\langle \mu \rangle \simeq \int d \ln k\, \Delta^2_\zeta(k)\, F(k)\, ,
\end{equation}
where $\Delta^2_\zeta(k)$ is the power spectrum of the primordial curvature perturbations. $F(k)$ is the $k$-space window function denoted as $W(k)$ in~\cite{Emami:2015xqa}: $F(k) \simeq (9/4) [e^{-2k^2/k^2_{D}(z_i)} - e^{-2k^2/k^2_{D}(z_f)} ]$
where $k_D(z)$  is the damping scale, and we need to evaluate the difference respectively at redshifts 
$z_i \sim 2 \times 10^{6}$ and $z_f \sim 5 \times 10^{4}$, defining the $\mu$-distortion era. Such redshifts correspond to diffusion scales $k_i \equiv k_D(z_i) \simeq 12000 \, {\rm Mpc}^{-1}$ and $k_f \equiv k_D(z_f) \simeq 46 \, {\rm Mpc}^{-1}$ \cite{1968ApJ...151..459S, 1970ApJ...162..815P, 1983MNRAS.202.1169K, Weinberg:2008zzc}. 
Let us now compute the $TT\mu$ bispectrum induced by the $g_{\rm NL}$ type trispectrum
\begin{eqnarray}
\label{main1}
\left \langle \frac{\delta T_1}{T} \frac{\delta T_2}{T} \frac{\delta \mu_3}{\mu} \right \rangle & \simeq & \frac{54}{25} g_{\rm NL} \left \langle \frac{\zeta_1}{5}  
\frac{\zeta_2}{5}   \left(\zeta^{\rm G}_{L3} \right)^2  \right \rangle  \nonumber \\ 
& =& 108\, g_{\rm NL}
\left \langle \frac{\delta T_1}{T} \frac{\delta T_3}{T} \right\rangle
\left \langle \frac{\delta T_2}{T} \frac{\delta T_3}{T} \right\rangle
\, . 
\end{eqnarray}
In Eq.~(\ref{main1}) the indices $1,2,3$ refer to three different positions on last-scattering surface (or, by means of an angular projection from the last-scattering surface, they label three different directions in the sky). Also, in writing Eq.~(\ref{main1}) we have used that the large-angle temperature fluctuation is given by $\delta T/T \simeq - \zeta/5$
(in the Sachs-Wolfe (SW) approximation). 

The equation above describes correlation between $\delta T/T$ and the fractional change in $\mu$-distortions, $\delta \mu/\mu$. If we want to 
work with $\mu$-fluctuations instead, we simply have to multiply Eq.~(\ref{main1}) by the average $\mu$ distortions, Eq.~(\ref{muaverage}). In the case of a scale invariant spectrum of primordial curvature perturbations with $\Delta^2_\zeta(k) = A_S$,  
$\langle \mu \rangle \simeq (9/4) A_S \ln (k_i/k_f)$, this yields  
\begin{equation}
\label{eqn:TTmu_gNL_intuitive}
b_{\ell_1 \ell_2 \ell_3}^{TT\mu}
\simeq 108\, g_{\rm NL}\, \frac{9}{4} A_S \ln\left(\frac{k_i}{k_f}\right) C_{\ell_1}^{TT} C_{\ell_2}^{TT}\, ,
\end{equation}
where we have moved to $\ell$ space by the harmonic transformation: $\left \langle \frac{\delta T_1}{T} \frac{\delta T_2}{T} \right\rangle \to C_{\ell_1}^{TT}$ and $\left \langle \frac{\delta T_1}{T} \frac{\delta T_2}{T} \delta \mu_3 \right\rangle \to b_{\ell_1 \ell_2 \ell_3}^{TT\mu}$, with $C_\ell^{TT}$ and $b_{\ell_1 \ell_2 \ell_3}^{TT\mu}$ denoting the angular power spectrum and bispectrum \eqref{eqn:factorization}, respectively.

The $TT\mu$ bispectrum induced by $\tau_{\rm NL}$-like NG can be computed in a similar way. 
Starting from Eq.~(\ref{eqn:taunl}), where the small-scale curvature perturbation $\zeta^{\rm G}({\bf x})$ is modulated by the large-scale field $\sigma({\bf x})$, 
we find (at leading order and up to disconnected parts)
\begin{equation}
\frac{\delta \langle \zeta^2 \rangle }{\langle \zeta^2\rangle }\simeq \frac{\delta \mu}{\mu}\simeq 2  \sqrt{\tau_{\rm NL}} \sigma({\bf x})\, .
\end{equation}
Notice that, following the conventional definition of $\tau_{\rm NL}$, in Eq.~(\ref{eqn:taunl}) $\sigma({\bf x})$ is normalized in such a way that it has equal power spectrum as $\zeta^{\rm G}({\bf x})$, $\langle \sigma^2 \rangle= \langle (\zeta^{\rm G})^2 \rangle$. Therefore 
\begin{eqnarray}
\label{main2}
\left \langle  \frac{\delta T_1}{T} \frac{\delta T_2}{T} \frac{\delta \mu_3}{\mu} \right \rangle
&\simeq& 
2 \left \langle \frac{\zeta_{1}^{\rm G}}{5}  
\frac{\zeta_{2}^{\rm G}}{5}
\left[ 1+\sqrt{\taunl} (\sigma_1 + \sigma_2) \right]
 \sqrt{\taunl} \sigma_3 \frac{}{}\right \rangle  
\nonumber \\ 
&=& 50 \taunl 
\left[\left\langle \frac{\delta T_1}{T} \frac{\delta T_2}{T}\right\rangle
\left\langle \frac{\delta T_2}{T} \frac{\delta T_3}{T}\right\rangle 
+(1 \leftrightarrow 2)\right]\, .
\end{eqnarray}
Finally, multiplying by the average $\mu$ distortion, we obtain the harmonic-space expression of $\left\langle  \frac{\delta T_1}{T} \frac{\delta T_2}{T} \delta \mu_3 \right\rangle$
\begin{equation}
b_{\ell_1 \ell_2 \ell_3}^{TT\mu} \simeq
50\, \tau_{\rm NL}\, \frac{9}{4} A_S \ln\left(\frac{k_i}{k_f}\right)
\left[ C_{\ell_1}^{TT} +C_{\ell_2}^{TT} \right] C_{\ell_3}^{TT} \,. \label{eqn:TTmu_tauNL_intuitive}
\end{equation}

In the next section, we will show how these results very nicely match a full detailed computation.

\section{The $TT\mu$ bispectrum} \label{sec:full}
  
After the warm up in the previous section, we are now ready to perform a full computation of the $TT\mu$ three-point function, arising from 
both $\gnl$ and $\taunl$ contributions. At the end of the section, we will find excellent agreement between the full and simplified treatments. 
Let us note, before starting our calculation, that $y$-type distortions could have been 
considered as well, and the $TTy$ bispectrum would produce contributions to the signal coming from a different range of scales. The authors of \cite{Pajer:2012vz} originally did not include y-contributions in their study of two-point correlations. This was based on the fact that primordial $Ty$ and $yy$ signals  would be affected by large contaminations coming from late-time Compton-y signals. It 
was however argued in \cite{Emami:2015xqa} that $yT$ primordial NG signatures could in principle be used to disentangle the high-redshift and low-redshift 
components. It remains anyway clear that $\mu$-T correlations provide the cleanest signal. We will thus focus here only on $TT\mu$, 
leaving issues related to $TTy$ contributions for future work.

CMB temperature anisotropies are linked, at first order, to primordial curvature perturbations via the usual formula:
\begin{eqnarray}
a_{\ell m}^{T} = 
4\pi i^{\ell} \int \frac{d^3 {\bf k}}{(2\pi)^{3}}
 {\cal T}_{\ell}(k) \zeta_{\bf k} Y_{\ell m}^*(\hat{\bf k}) \; ,
\end{eqnarray}
where  ${\cal T}_{\ell}(k)$ indicates the radiation transfer function, and $\zeta_{\bf k}$ is the primordial curvature perturbation. 

The $\mu$ spectral distortion parameter from dissipation of acoustic fluctuations can instead be obtained as 
(e.g. \cite{Sunyaev:1970er, 1975SvA....18..413I, 1982A&A...107...39D, 1991A&A...246...49B, Hu:1994bz, Chluba:2011hw, Khatri:2011aj, Chluba:2012gq, Khatri:2012tv, Khatri:2012tw,1968ApJ...151..459S, 1970ApJ...162..815P, 1983MNRAS.202.1169K,Weinberg:2008zzc,Pajer:2012vz, Ganc:2012ae, Khatri:2015tla}):
\begin{equation}
a_{\ell m}^\mu = 
4\pi (-i)^{\ell} \left[ \prod_{n=1}^2 
\int \frac{d^3 {\bf k}_n}{(2\pi)^3} \zeta_{{\bf k}_n}
\right] 
\int d^3 {\bf k}_3 
\delta^{(3)}\left({\bf k}_1 + {\bf k}_2 + {\bf k}_3 \right) Y_{\ell m}^*(\hat{\bf k}_3)
 j_{\ell}(k_3 x_{\rm ls}) f(k_1, k_2, k_3) \; , 
\end{equation}
where $x_{\rm ls}$ is
the conformal distance to the last scattering surface
and:
\begin{equation}
f(k_1, k_2, k_3) \simeq \frac{9}{4} W\left(\frac{k_3}{k_s}\right) \left[e^{-(k_1^2+k_2^2)/k_D^2(z)}\right]_{f}^{i} \; .
\end{equation}
In the last formula, $W(k/k_s)$ is a window function selecting the range of scales $k/k_s \lesssim 1$ for acoustic wave dissipation, and the square bracket takes the difference between the quantities at $z_i \sim 2 \times 10^{6}$ and $z_f \sim 5 \times 10^{4}$, namely, $[ g(z) ]_{f}^{i} \equiv g(z_i) - g(z_f) $. The $\langle a_{\ell_1 m_1}^{T} a_{\ell_2 m_2}^{T} a_{\ell_3 m_3}^\mu \rangle$ bispectrum can now be written as:
\begin{eqnarray}
  \Braket{a_{\ell_1 m_1}^T a_{\ell_2 m_2}^T a_{\ell_3 m_3}^\mu}
  &=& 
 \left[\prod_{n=1}^2 4\pi i^{\ell_n} \int \frac{d^3 {\bf k}_n}{(2\pi)^{3}}
  {\cal T}_{\ell_n}(k_n)  Y_{\ell_n m_n}^*(\hat{\bf k}_n) \right] \nonumber \\ 
  && 4\pi (-i)^{\ell_3} \left[ \prod_{n=1}^2 
\int \frac{d^3 {\bf K}_n}{(2\pi)^3} 
\right]
\int d^3 {\bf K}_3 \delta^{(3)}\left({\bf K}_1 + {\bf K}_2 + {\bf K}_3 \right) \nonumber \\
&& Y_{\ell_3 m_3}^*(\hat{\bf K}_3)  j_{\ell_3}(K_3 x_{\rm ls}) f(K_1, K_2, K_3)
\Braket{ \zeta_{{\bf k}_1} \zeta_{{\bf k}_2}  \zeta_{{\bf K}_1}  \zeta_{{\bf K}_2}} \, .
\label{eqn:TTm} 
\end{eqnarray}

\subsection{$\gnl$ contributions}

The $\gnl$-type trispectrum is given as
\begin{eqnarray}
\Braket{ \zeta_{{\bf k}_1} \zeta_{{\bf k}_2}  \zeta_{{\bf K}_1}  \zeta_{{\bf K}_2} }
   =  (2\pi)^3 \delta^{(3)}\left({\bf k}_1 + {\bf k}_2 + {\bf K}_1 + {\bf K}_2 \right) 
T(k_1, k_2 ,K_1, K_2) \; ,
\end{eqnarray}
 with
\begin{equation}
T(k_1, k_2 ,K_1, K_2) = \frac{54}{25} \gnl \left[ P(k_1) P(k_2) P(K_1) + (3~{\rm perm})\right]~. \label{eqn:zeta4_gNL}
\end{equation}
Substituting this into Eq.~\eqref{eqn:TTm},
it is possible to arrive at the following expression
\begin{eqnarray}
  \Braket{a_{\ell_1 m_1}^T a_{\ell_2 m_2}^T a_{\ell_3 m_3}^\mu} &=&
 \left[\prod_{n=1}^2  4\pi i^{\ell_n} \int \frac{d^3 {\bf k}_n}{(2\pi)^{3}} {\cal T}_{\ell_n}(k_n)  Y_{\ell_n m_n}^*(\hat{\bf k}_n) \right] \nonumber \\ 
 &&
 4\pi (-i)^{\ell_3} \left[ \prod_{n=1}^2 
\int \frac{d^3 {\bf K}_n}{(2\pi)^3} 
\right]
         \int d^3 {\bf K}_3
         \nonumber \\
         &&  8 \int_0^\infty y^2 dy 
\left[ \prod_{n=1}^3 \sum_{L_n M_n} 
 j_{L_n}(K_n y) 
 Y_{L_n M_n}^*(\hat{\bf K}_n) \right] \nonumber \\ 
 &&
(-1)^{\frac{L_1 + L_2 + L_3}{2}}
h_{L_1 L_2 L_3}
\left(
  \begin{array}{ccc}
  L_1 & L_2 & L_3 \\
  M_1 & M_2 & M_3 
  \end{array}
 \right) 
\nonumber \\
&& 
Y_{\ell_3 m_3}^*(\hat{\bf K}_3)
j_{\ell_3}(K_3 x_{\rm ls}) f(K_1, K_2, K_3)
(2\pi)^3 T(k_1, k_2, K_1, K_2)
\nonumber \\
&& 
2^5 \pi \int_0^\infty r^2 dr
\left[ \prod_{n=1}^2 \sum_{l_n' m_n'}
j_{l_n'}(k_n r) Y_{l_n' m_n'}^*(\hat{\bf k}_n)
 \sum_{L_n' M_n'}
j_{L_n'}(K_n r) Y_{L_n' M_n'}^*(\hat{\bf K}_n)
  \right] \nonumber \\ 
 &&
(-1)^{\frac{l_1' + l_2' + L_1' + L_2'}{2}}
\sum_{L' M'} (-1)^{M'} h_{l_1' l_2' L'} h_{L_1' L_2' L'} \nonumber \\ 
 &&
  \left(
  \begin{array}{ccc}
  l_1' & l_2' & L' \\
  m_1' & m_2' & M' 
  \end{array}
  \right)
  \left(
  \begin{array}{ccc}
  L_1' & L_2' & L' \\
  M_1' & M_2' & -M' 
  \end{array}
 \right) 
 \label{eqn:TTm2_gNL} \; ,
\end{eqnarray}
where $ \left( \begin{array}{ccc}
  \ell_1 & \ell_2 & \ell_3 \\
  m_1 & m_2 & m_3 
  \end{array} \right) $ are Wigner-3j symbols, and we defined:
\begin{equation}
h_{l_1 l_2 l_3} 
\equiv \sqrt{\frac{(2 l_1 + 1)(2 l_2 + 1)(2 l_3 + 1)}{4 \pi}}
\left(
  \begin{array}{ccc}
  l_1 & l_2 & l_3 \\
   0 & 0 & 0 
  \end{array}
 \right)  \; . \\
\end{equation}

In order to go from Eq.~(\ref{eqn:TTm}) to Eq.~(\ref{eqn:TTm2_gNL}) we used integral representations for the Dirac delta functions, expanding plane waves in spherical harmonics:
\begin{equation}
  e^{i{\bf k} \cdot {\bf x}} = \sum_{L M}
  4\pi i^{L}   j_L(kx) Y_{LM}(\hat{\bf k}) Y_{LM}^*(\hat{\bf x}) \; ,
\end{equation}
and used the Gaunt integral representation for integrals of products of three spherical harmonics. We can now use the orthonormality relation for spherical harmonics:
\begin{equation}
\int d^2 \hat{\bf n} Y_{\ell_1 m_1}(\hat{\bf n}) Y_{\ell_2 m_2}^*(\hat{\bf n}) = \delta_{\ell_1, \ell_2} \delta_{m_1, m_2} \; ,
\end{equation}
and the completeness of Wigner-3j symbols:
\begin{equation}
\frac{\delta_{l_3, l_3'} \delta_{m_3, m_3'}}{(2l_3+1)}
  = \sum_{m_1 m_2}
  \left(
  \begin{array}{ccc}
  l_1 & l_2 & l_3 \\
  m_1 & m_2 & m_3
  \end{array}
  \right)
  \left(
  \begin{array}{ccc}
  l_1 & l_2 & l_3' \\
  m_1 & m_2 & m_3' 
  \end{array}
  \right)  \; ,
\end{equation}
to further simplify Eq.~\eqref{eqn:TTm2_gNL} into:
\begin{equation}\label{eqn:factorization}
\Braket{a_{\ell_1 m_1}^T a_{\ell_2 m_2}^T a_{\ell_3 m_3}^\mu}=  h_{\ell_1 \ell_2 \ell_3}
  \left(
  \begin{array}{ccc}
  \ell_1 & \ell_2 & \ell_3 \\
  m_1 & m_2 & m_3 
  \end{array}
  \right)
  b_{\ell_1 \ell_2 \ell_3}^{TT\mu} \; ,
\end{equation}
where we have defined the {\em reduced bispectrum}:
\begin{eqnarray} 
b_{\ell_1 \ell_2 \ell_3}^{TT\mu} & = & 
\int_0^\infty r^2 dr \int_0^\infty y^2 dy \,
 \frac{2}{\pi} \int_0^\infty k_1^2 d k_1 {\cal T}_{\ell_1}(k_1) j_{\ell_1}(k_1 r) \,
 \frac{2}{\pi} \int_0^\infty k_2^2 d k_2 {\cal T}_{\ell_2}(k_2) j_{\ell_2}(k_2 r) \,
  \nonumber \\
  & &
\sum_{L_1 L_2} \frac{h_{L_1 L_2 \ell_3}^2}{2\ell_3 + 1} 
  \frac{2}{\pi} \int_0^\infty K_1^2 dK_1 j_{L_1}(K_1 y) j_{L_1}(K_1 r) \, 
  \frac{2}{\pi} \int_0^\infty K_2^2 dK_2 j_{L_2}(K_2 y) j_{L_2}(K_2 r) \,
  \nonumber \\
&& 
\frac{2}{\pi} \int_0^\infty K_3^2 d K_3
j_{\ell_3}(K_3 x_{\rm ls}) j_{\ell_3}(K_3 y) 
f(K_1, K_2, K_3) T(k_1, k_2, K_1, K_2) \; . \label{eqn:reduced_gNL}
\end{eqnarray}
The factorization in Eq.~(\ref{eqn:factorization}) is, as usual in this type of calculations (see e.g.,~\cite{Komatsu:2001rj}), a direct consequence of the rotational invariance properties of the CMB sky. All physical information is contained in the reduced bispectrum defined in Eq.~(\ref{eqn:reduced_gNL}). 
Up to this point, we performed an exact calculation. 
In order to make Eq.~(\ref{eqn:reduced_gNL}) feasible for numerical evaluation, we now simplify it by using the following approximation:
\begin{eqnarray}
 \int_0^\infty K_3^2 d K_3
 j_{\ell_3}(K_3 x_{\rm ls}) j_{\ell_3}(K_3 y) W\left(\frac{K_3}{k_s}\right)
 &\simeq&
 \int_0^\infty K_3^2 d K_3
 j_{\ell_3}(K_3 x_{\rm ls}) j_{\ell_3}(K_3 y) \nonumber \\
 &=& 
  \frac{\pi \delta(y - x_{\rm ls})}{2 x_{\rm ls}^2}~. \label{eqn:int_K3}
\end{eqnarray}
This is justified by the fact that the spherical Bessel function $j_{\ell}(kx)$ is peaked for $\ell \sim kx$, and decays rapidly afterwards. We can regard $k_s$ in the argument of the window function, $W(K_3/k_s)$, as the damping scale at recombination $k_D(z_* \simeq 1100)$ \cite{Khatri:2015tla}, and the integral then converges to very high accuracy well before the $K_3 \sim k_D(z_*)$ cutoff, as long as $\ell_3 \ll k_D(z_*) x_{\rm ls} \simeq 2000$. 
This condition will always be verified in the following, since in our forecasts we will take $\ell_1 = \ell_2 = \ell_3 = 1000$ as our maximum value. The last equality in Eq.~\eqref{eqn:int_K3} expresses the completeness of 
spherical Bessel functions.
Plugging Eq.~(\ref{eqn:int_K3}) and the $\gnl$ trispectrum formula \eqref{eqn:zeta4_gNL} into Eq.~(\ref{eqn:reduced_gNL}) finally yields:
  \begin{eqnarray}    
  b_{\ell_1 \ell_2 \ell_3}^{TT\mu}
 &\simeq&  
  \frac{54}{25} \gnl
  \sum_{L_1 L_2} \frac{h_{L_1 L_2 \ell_3}^2}{2\ell_3 + 1}
   \int_0^\infty r^2 dr 
\nonumber \\
&& 
\left[
  \beta_{\ell_1}^T(r) \beta_{\ell_2}^T(r) \beta_{L_1}^\mu(r,z) \alpha_{L_2}^\mu(r,z) 
  + \beta_{\ell_1}^T(r) \beta_{\ell_2}^T(r) \alpha_{L_1}^\mu(r,z) \beta_{L_2}^\mu(r,z) \nonumber \right. \\ 
  &&\left. + \beta_{\ell_1}^T(r) \alpha_{\ell_2}^T(r)\beta_{L_1}^\mu(r,z) \beta_{L_2}^\mu(r,z) 
  + \alpha_{\ell_1}^T(r) \beta_{\ell_2}^T(r)  \beta_{L_1}^\mu(r,z) \beta_{L_2}^\mu(r,z)
 \right]_{f}^{i}\label{eqn:redgnl} ~,
  \end{eqnarray}
where we have defined:
\begin{eqnarray}
  \alpha_{\ell}^T(r)
  &\equiv& \frac{2}{\pi} \int_0^\infty k^2 d k {\cal T}_\ell(k) j_{\ell}(k r) ~, \\
  \alpha_{\ell}^\mu(r,z)
  &\equiv& \frac{3}{\pi} \int_0^\infty k^2 d k j_{\ell}(k x_{\rm ls}) j_{\ell}(k r)
   e^{-k^2/k_D^2(z)}\label{eqn:alphamu} ~,  \\
  \beta_{\ell}^T(r)
  &\equiv& \frac{2}{\pi} \int_0^\infty k^2 d k P(k) {\cal T}_{\ell}(k) j_{\ell}(k r) ~, \\
  \beta_{\ell}^\mu(r,z)
  &\equiv& \frac{3}{\pi} \int_0^\infty k^2 d k P(k) j_{\ell}(k x_{\rm ls}) j_{\ell}(k r)
  e^{-k^2/k_D^2(z)}. \label{eqn:betamu}
  \end{eqnarray}

\begin{figure}[t!]
  \begin{tabular}{c}
    \begin{minipage}{1.0\hsize}
  \begin{center}
    \includegraphics[width = 0.8\textwidth]{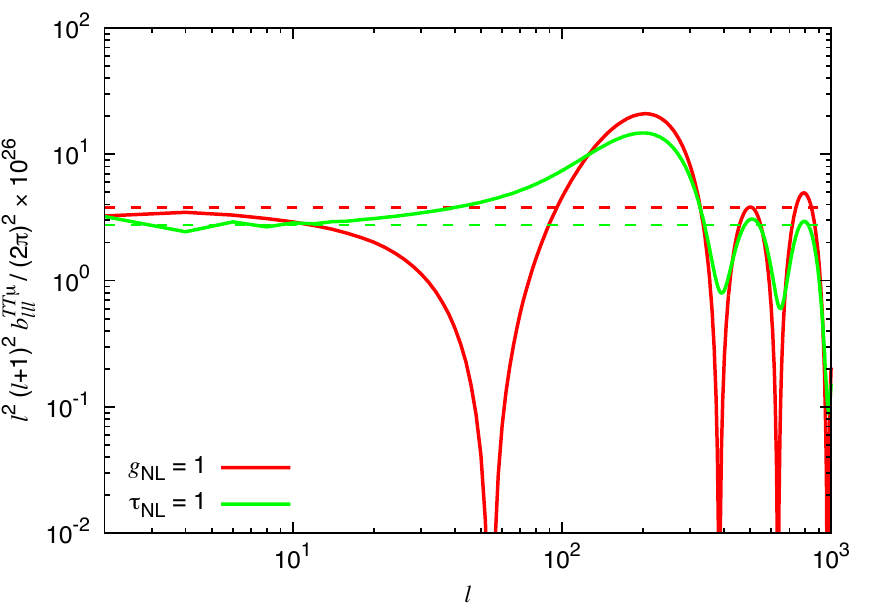}
  \end{center}
\end{minipage}
\end{tabular}
  \caption{$TT\mu$ for $\gnl = 1$ and $\taunl = 1$. The solid and dashed lines describe the results including full transfer function (computed from Eqs.~\eqref{eqn:redgnl_fisher} and \eqref{eqn:redtau_fisher}) and those in the SW limit (computed from Eqs.~\eqref{eqn:TTmu_gNL_SW_approx} and \eqref{eqn:TTmu_tau_SW_approx}), respectively. As expected, the SW approximation agrees well with the full calculation on very small $\ell$'s.}\label{fig:blll_TTmu} 
\end{figure}

Let us now consider a scale-invariant primordial power spectrum, $P(k) = 2\pi^2 A_S k^{-3}$. Using again asymptotic properties and the completeness 
relation for spherical Bessel functions, as well as keeping into account the $k_D$ cutoff in Eqs.~(\ref{eqn:alphamu}) and (\ref{eqn:betamu}), 
we can approximately evaluate $\alpha_\ell^\mu$ and $\beta_\ell^\mu$ as follows:
\begin{eqnarray} 
  \alpha_L^\mu(r,z) &\approx&
\begin{cases}
  \frac{3\delta(r - x_{\rm ls})}{2x_{\rm ls}^2} &: L \lesssim L_D(z) \\
 0 &: L \gtrsim L_D(z)
\end{cases} ~, \label{eqn:alpha_final} \\
\beta_L^\mu(x_{\rm ls},z) &\approx&
\begin{cases}
 \frac{3 \pi A_S}{L(L+1)} &: L \lesssim L_D(z) \\
 0 &: L \gtrsim L_D(z)
\end{cases} \label{eqn:beta_final} \; ,
\end{eqnarray}
 where we have defined $L_D(z) \equiv k_D(z) x_{\rm ls}$. 
In Eq.~(\ref{eqn:redgnl}) we can thus operate the replacement 
\begin{equation}
\sum_{L_1, L_2=0}^{\infty} \rightarrow 
\sum_{L_1, L_2 = L_f}^{L_i} \; ,
\end{equation}
with $L_f \equiv k_f x_{\rm ls} \sim 10^5$ and $L_i \equiv k_i x_{\rm ls} \sim 10^8$. We can then see that $L_1, L_2$ are very large 
and $L_1,L_2 \gg \ell_3$ in the sum. In light of this, and using the Stirling approximation to evaluate the Wigner symbols, we can get the following asymptotic 
formula:
\begin{eqnarray}
  h_{l_1 l_2 l_3}^2
  \simeq \frac{2 l_1}{\pi^2}\label{eqn:hl1l2l3sq_approx} \; ,
  \end{eqnarray}
where we assume $l_1 \simeq l_2$, in virtue of the triangle inequality, imposing $|l_1 - l_2| \leq l_3 \leq l_1 + l_2$, and of the condition $l_1, l_2 \gg l_3$.

We tested the approximation \eqref{eqn:hl1l2l3sq_approx} by comparing $h_{l_1 l_2 l_3}^2$ computed numerically with $2 l_1 / \pi^2$, for $l_1 \sim 1000$, and found that the error it introduces is $\sim 20 \%$, which is 
completely reasonable for the order of magnitude Fisher forecasts in the next section. We can then operate the further replacement:
\begin{equation}
\sum_{L_1, L_2 = L_f}^{L_i} \frac{h_{L_1 L_2 \ell_3}^2}{2\ell_3 + 1}
  \simeq \sum_{L_1,L_2 = L_f}^{L_i} \frac{2 L_1}{\pi^2} \delta_{L_1, L_2}  \; .
\end{equation}
After substituting Eqs.~(\ref{eqn:alpha_final}) and (\ref{eqn:beta_final}) into Eq.~(\ref{eqn:redgnl}), keeping the leading order terms in the sum over $L_1$, and evaluating the sum via integration, we finally arrive at the expression:
\begin{equation} 
  b_{\ell_1 \ell_2 \ell_3}^{TT\mu}
 \simeq 
 \frac{972}{25\pi} \gnl A_S 
 \ln\left(\frac{k_i}{k_f}\right)
 \beta_{\ell_1}^T(x_{\rm ls}) \beta_{\ell_2}^T(x_{\rm ls}) ~.
\label{eqn:redgnl_fisher}
\end{equation}
If we take the Sachs-Wolfe (SW) limit, ${\cal T}_\ell(k) \to -\frac{1}{5}j_\ell(k x_{\rm ls}) $, we can further simplify this into the following analytical expression:
\begin{equation}
    b_{\ell_1 \ell_2 \ell_3}^{TT\mu, \rm SW}
  \simeq \frac{972}{\pi} \gnl  A_S
  \ln\left(\frac{k_i}{k_f}\right)
  C_{\ell_1, \rm SW}^{TT} C_{\ell_2, \rm SW}^{TT} \label{eqn:TTmu_gNL_SW_approx} \; ,
\end{equation}
where $C_{\ell, \rm SW}^{TT} = \frac{2\pi A_S}{25 \ell (\ell+1)}$. This is consistent with our expectation in Eq.~\eqref{eqn:TTmu_gNL_intuitive} (the $4/\pi$ difference is simply due to the approximation in Eq.~\eqref{eqn:hl1l2l3sq_approx}).

\subsection{$\taunl$ contributions}

The $TT\mu$ signal arising from a primordial local $\taunl$-trispectrum can be computed in similar fashion as we did for the $\gnl$ part. Starting 
from Eq.~(\ref{eqn:TTm}), we include the $\taunl$ shape:
\begin{equation} 
 \Braket{\zeta_{{\bf k}_1} \zeta_{{\bf k}_2} \zeta_{{\bf K}_1}  \zeta_{{\bf K}_2}}
= (2\pi)^3 \delta^{(3)}\left({\bf k}_1 + {\bf k}_2 + {\bf K}_1 + {\bf K}_2 \right) 
\taunl 
\left[ P(k_1) P(K_1) P(k_{12}) + (11~{\rm perm}) \right] ~, 
\end{equation}
where $k_{12} \equiv |{\bf k}_1 + {\bf k}_2|$.
We then take into account the fact that the filters $f(K_1,K_2,K_3)$ select configurations for which 
$K_1 \simeq K_2 \gg K_{12}$. This allows to write:
\begin{eqnarray}
\Braket{\zeta_{{\bf k}_1} \zeta_{{\bf k}_2} \zeta_{{\bf K}_1}  \zeta_{{\bf K}_2}}
&\simeq& (2\pi)^3 \int d^3 {\bf k} \delta^{(3)}\left({\bf k}_1 + {\bf k}_2 + {\bf k} \right) 
\delta^{(3)}\left({\bf K}_1 + {\bf K}_2 - {\bf k} \right) t_{K_1 K_2}^{k_1 k_2}(k) ~, \\ 
t_{K_1 K_2}^{k_1 k_2}(k) &\equiv& \taunl \left[ P(k_1) P(K_1) + P(k_1) P(K_2) 
 + P(k_2) P(K_1)  + P(k_2) P(K_2)  \right] P(k)~. \nonumber  
\end{eqnarray}
Using the expansions and properties which lead from Eq.~(\ref{eqn:TTm2_gNL}) to Eq.~(\ref{eqn:reduced_gNL}), we arrive at:
\begin{eqnarray}
b_{\ell_1 \ell_2 \ell_3}^{TT\mu}
  &=& \frac{8}{\pi^6}
\int_0^\infty r^2 dr 
 \int_0^\infty k_1^2 dk_1 {\cal T}_{\ell_1}(k_1) j_{\ell_1}(k_1 r)
 \int_0^\infty k_2^2 dk_2 {\cal T}_{\ell_2}(k_2) j_{\ell_2}(k_2 r)  \nonumber \\
 && 
 \int_0^\infty K_1^2 d K_1
\int_0^\infty K_2^2 dK_2 
\int_0^\infty K_3^2 dK_3 j_{\ell_3}(K_3 x_{\rm ls}) j_{\ell_3}(K_3 r) \nonumber \\
 && 
\int_0^\infty y^2 dy  j_{0}(K_1 y) j_{0}(K_2 y) j_{0}(K_3 y)
f(K_1, K_2, K_3) t_{K_1 K_2}^{k_1 k_2}(K_3) \; .
\end{eqnarray}
In the limit $K_1 \sim K_2 \gg K_3$, we have $j_0(K_3 y) \rightarrow 1$, and:
\begin{equation}
  \int_0^\infty y^2 dy  j_{0}(K_1 y) j_{0}(K_2 y) j_{0}(K_3 y)
  \simeq \frac{\pi \delta(K_1 - K_2)}{2 K_1^2} ~.
  \end{equation}
With this approximation we can write 
\begin{eqnarray}
  b_{\ell_1 \ell_2 \ell_3}^{TT\mu} &\simeq& \frac{9}{2}\taunl A_S \ln \left( \frac{k_i}{k_f} \right) 
  \int_0^\infty r^2 dr \left[ \alpha_{\ell_1}^T(r) \beta_{\ell_2}^T(r) + \beta_{\ell_1}^T(r) \alpha_{\ell_2}^T(r) \right]  
 \omega_{\ell_3}(r) \label{eqn:redtau_fisher} ~,
\end{eqnarray}
where
\begin{eqnarray}
 \omega_\ell(r) \equiv \frac{2}{\pi} \int_0^\infty  k^2 dk P(k)  j_{\ell}(k x_{\rm ls}) j_{\ell}(k r)
 W\left(\frac{k}{k_s}\right).
\end{eqnarray}
In the SW limit for $\alpha_\ell^T$ and $\beta_\ell^T$, and approximating $\omega_{\ell_3}(x_{\rm ls})$, for $\ell_3 \ll k_s x_{\rm ls} \simeq 2000$, in the following way:
\begin{equation}
  \omega_{\ell_3}(x_{\rm ls})
  \simeq \frac{2}{\pi} \int_0^\infty  k^2 dk P(k)  j_{\ell_3}^2 (k x_{\rm ls}) 
  =   \frac{2\pi A_S}{\ell_3(\ell_3 + 1)}
  ~,
  \end{equation}
we arrive at an analytical expression for the $TT\mu$ bispectrum originated by a primordial $\taunl$-signal
\begin{eqnarray}
  b_{\ell_1 \ell_2 \ell_3}^{TT\mu, \rm SW} =  \frac{225}{2} \taunl A_S 
  \ln \left( \frac{k_i}{k_f} \right) 
  \left[ C_{\ell_1, \rm SW}^{TT} + C_{\ell_2, \rm SW}^{TT} \right] C_{\ell_3, \rm SW}^{TT}~, \label{eqn:TTmu_tau_SW_approx}
\end{eqnarray}
which is completely consistent with our intuitive estimation \eqref{eqn:TTmu_tauNL_intuitive}. Formulae (\ref{eqn:redgnl_fisher}), (\ref{eqn:TTmu_gNL_SW_approx}), (\ref{eqn:redtau_fisher}) and (\ref{eqn:TTmu_tau_SW_approx}) will be our starting point both for numerical evaluation of the $TT\mu$ $\gnl$-and $\taunl$-bispectra, which are displayed in Fig.~\ref{fig:blll_TTmu}, and for Fisher forecasting in the next section.

\subsection{Contributions of the Gaussian part} 

Before concluding this section it is however important to consider whether Gaussian contributions to the trispectrum might produce 
a bias in $\gnl$ and $\taunl$ measurements from the $TT\mu$ signal. The short answer is ``no'', and this is due again to the fact that $\mu$-distortions 
 filters very small scales, while temperature anisotropies are generated at large scales, so that, in absence of mechanisms coupling short and long modes, the two are uncorrelated. A full calculation confirms this. We start with the primordial 4-point function 
generated by Gaussian primordial perturbations. If we neglect disconnected term, contributing only to the monopole, this reads
\begin{eqnarray}
    \Braket{\zeta_{{\bf k}_1} \zeta_{{\bf k}_2} \zeta_{{\bf K}_1} \zeta_{{\bf K}_2}  }
    =  (2\pi)^6  P(k_1) P(k_2)
\delta^{(3)}\left({\bf k}_1 + {\bf K}_1 \right) 
\delta^{(3)}\left( {\bf k}_2 + {\bf K}_2 \right)
+ ({\bf k}_1 \leftrightarrow {\bf k}_2 )  ~.
  \end{eqnarray}
If we plug this into Eq.~(\ref{eqn:TTm}), and follow analogous steps as for the calculation of the $\gnl$ signal, we obtain, keeping into account 
the approximation used in Eq.~(\ref{eqn:int_K3}) :
 \begin{eqnarray}
  b_{\ell_1 \ell_2 \ell_3}^{TT\mu}
  \simeq 2 \left[ \tilde{\beta}_{\ell_1}^T(z) \tilde{\beta}_{\ell_2}^T(z) \right]_f^i \; ,
 \end{eqnarray}
   with
   \begin{equation}
  \tilde{\beta}_\ell^T(z) \equiv \frac{3}{\pi} \int_0^\infty k^2 dk P(k) {\cal T}_{\ell}(k) j_{\ell}(k x_{\rm ls}) e^{-k^2/k_D^2(z)} ~.
   \end{equation}
   In the SW limit, we can take $\tilde{\beta}_\ell^T(z) \to -\frac{1}{5} \beta_\ell^\mu(x_{\rm ls}, z)$. If we now consider the asymptotic approximation (\ref{eqn:beta_final}), we can see how this quantity is essentially vanishing in the relevant range of 
scales $\ell_1, \ell_2, \ell_3 \ll L_f$. Assuming Gaussianity of the noise, for a given experiment, we can then conclude that the $TT\mu$ statistic is able 
to provide {\em unbiased} estimates of the local trispectrum parameters $\gnl$ and $\taunl$.

\section{Forecasts} \label{sec:fisher}

If we consider a case with $\fnl=0$, we can forecast error bars on $TT\mu$ estimates of $\gnl$ and $\taunl$ using the 
Fisher matrix:
\begin{eqnarray}
 F^{TT\mu} = \sum_{\ell_1 \ell_2 \ell_3} \frac{\left(h_{\ell_1 \ell_2 \ell_3} \hat{b}_{\ell_1 \ell_2 \ell_3}^{TT\mu} \right)^2}{2 C_{\ell_1}^{TT} C_{\ell_2}^{TT} C_{\ell_3}^{\mu\mu} } \;, \label{eqn:Fisher}
\end{eqnarray}
where $\hat{b}^{TT\mu}$ denotes the $TT\mu$ bispectrum normalized at $\gnl = 1$ or $\taunl = 1$, and we took $C_\ell^{T\mu}  = 0$ in the denominator, as it is the case 
when $\fnl=0$. Regarding the $\mu \mu$ contribution to the denominator, the contribution arising from the Gaussian part of the signal is computed as $C_\ell^{\mu\mu, \rm G} \sim 10^{-30}$ for $\ell \lesssim 1000$, in the same manner as \cite{Pajer:2012vz}. We note here that, if $\taunl$ does not 
vanish, the NG contribution to $\mu\mu$ dominates over the Gaussian part at small $\ell$'s (the G contribution is constant, while the NG part scales 
like $\ell^{-2}$ \cite{Pajer:2012vz}).
We account for the degradation of the error bars, obtained with the inclusion of this NG contribution, by simply adding it
to $C_\ell^{\mu\mu}$ in the denominator 
of Eq.~(\ref{eqn:Fisher}). A full forecast, including different fiducial values of $\fnl$, $\gnl$ and $\taunl$ and the joint covariance between $2$  
and $3$-point signals, while interesting, is beyond the scope of the current analysis, and will be pursued in future work. Regarding the contribution to $\mu \mu$  
arising from the $\gnl$-part of the primordial trispectrum, similar calculations to those performed in \cite{Pajer:2012vz} for the $\taunl$-part show that this is negligible with respect to the Gaussian part, for values of $\gnl$ which are not ruled out by $\Planck$ \cite{Ade:2015ava}. We find in fact $C_\ell^{\mu\mu, \gnl} \sim 10^{-37} \gnl$ for $\ell \lesssim 1000$.

\subsection{Cosmic-variance dominated measurements}

\begin{figure}
\begin{tabular}{c}
    \begin{minipage}{1.0\hsize}
  \begin{center}
    \includegraphics[width = 0.8\textwidth]{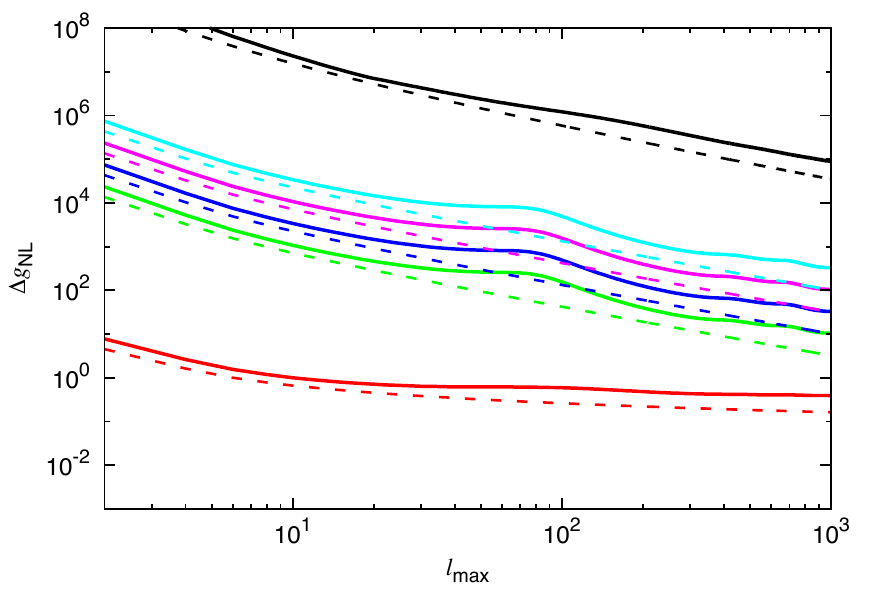}
  \end{center}
\end{minipage}
\end{tabular}
\\
\begin{tabular}{c}
    \begin{minipage}{1.0\hsize}
  \begin{center}
    \includegraphics[width = 0.8\textwidth]{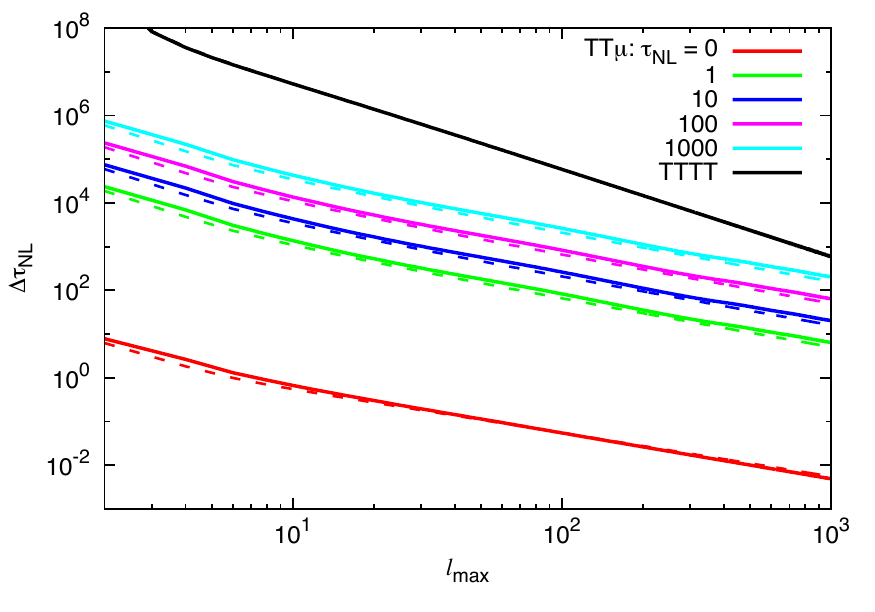}
  \end{center}
\end{minipage}
\end{tabular}
\caption{Expected 1$\sigma$ errors on $\gnl$ (top panel) and $\taunl$ (bottom panel) estimated from $TT\mu$ (colored lines) and $TTTT$ (black lines) in the cosmic-variance dominated case (i.e., $N_\ell^{\mu \mu} = 0$). Solid and dashed lines are the full radiation transfer case (Eqs.~\eqref{eqn:redgnl_fisher} and \eqref{eqn:redtau_fisher} for $TT\mu$) and the SW case (Eqs.~\eqref{eqn:TTmu_gNL_SW_approx} and \eqref{eqn:TTmu_tau_SW_approx} for $TT\mu$), respectively. In the $TT\mu$ cases, we consider several nonzero $\taunl$'s with $\fnl = 0$. For $\taunl = 0$, $\Delta \gnl$ and $\Delta \taunl$ obtained from $TT\mu$ scale like $1/\ln(\ell_{\rm max} / 2)$ and $1/\ell_{\rm max}$, respectively (see Eqs.~\eqref{eqn:dgNL_tau0} and \eqref{eqn:dtauNL_tau0}). It is apparent that, if $\taunl \leq 1000$, for $\ell_{\rm max} \leq 1000$, $TT\mu$ always outperforms $TTTT$, because $C_\ell^{\mu\mu, \rm G} + C_\ell^{\mu\mu, \taunl} \ll C_\ell^{TT}$. At larger $\ell_{\rm max}$, $TT\mu$ remains clearly superior to $TTTT$ for $\gnl$ measurements. For $\taunl$ estimation the comparison is instead dependent on the fiducial value of $\taunl$; see main text for further discussion.} \label{fig:dgNL_dtauNL_CV}
\end{figure}

The expected $1\sigma$ errors on $\gnl$ and $\taunl$, given by $\Delta \gnl, \Delta \taunl = 1/\sqrt{F^{TT\mu}}$, in the cosmic variance dominated regime are shown in Fig.~\ref{fig:dgNL_dtauNL_CV}. For a futuristic, cosmic variance dominated experiment up to $\ell \sim 1000$ (in $\mu$), we can see that spectral distortion based estimators can produce extremely tight error bars, $\Delta \gnl \simeq 0.4$ and $\Delta \taunl \simeq 5 \times 10^{-3}$, for fiducial values $\fnl = 0$, and $\taunl = 0$. This, as originally pointed out in \cite{Pajer:2012vz}, is due to the fact that $\mu$ spectral distortions provide (integrated) information up to very high wavenumbers. However, if we have $\taunl \neq 0$, the sensitivity is reduced due to the increase of $C_\ell^{\mu\mu}$, as described in Fig.~\ref{fig:dgNL_dtauNL_CV}.  This in particular implies that $\taunl = 5\times 10^{-3}$ is {\em not} the smallest detectable $\taunl$, since the error bar computed for this central value satisfies $\Delta \taunl|_{\taunl = 5\times 10^{-3}} > 5\times 10^{-3}$. An inspection of the bottom panel of Fig.~\ref{fig:dgNL_dtauNL_CV} and a fact that
 $\Delta \taunl$ scales like $\sqrt{\taunl}$ for $\taunl \gtrsim 1$ and $\ell \lesssim 1000$ (due to $C_\ell^{\mu\mu, \taunl} \sim 5 \times 10^{-23} \taunl \ell^{-2} \gg C_\ell^{\mu\mu, \rm G}$) shows that the smallest value of the parameter for which $\Delta \taunl|_{\taunl = {\bar{\tau}}_{\rm NL} }< \bar{\tau}_{\rm NL}$ (at $\ell_{\rm max} = 1000$) corresponds to $\bar{\tau}_{\rm NL} \sim 40$.

To understand the $\ell_{\rm max}$ dependence of $\Delta \gnl$ and $\Delta \taunl$, we can estimate Eq.~\eqref{eqn:Fisher} analytically, using the flat-sky approximation \cite{Hu:2000ee, Babich:2004yc}
\begin{eqnarray}
  F^{TT\mu}
  \simeq \frac{1}{\pi(2\pi)^2} \left[\prod_{n=1}^3 \int d^2 \boldsymbol{\ell}_n \right]  
  \delta^{(2)}\left( \boldsymbol{\ell}_1 + \boldsymbol{\ell}_2 + \boldsymbol{\ell}_3  \right) \frac{\left(\hat{b}_{\ell_1 \ell_2 \ell_3}^{TT\mu} \right)^2}{2 C_{\ell_1}^{TT} C_{\ell_2}^{TT} C_{\ell_3}^{\mu\mu} }~.
\end{eqnarray}
This should be accurate for large $\ell$. For simplicity, we work here with the SW formulae \eqref{eqn:TTmu_gNL_SW_approx} and \eqref{eqn:TTmu_tau_SW_approx} and assume $\tau_{\rm NL} = 0$, i.e., $C_\ell^{\mu\mu} = C_\ell^{\mu\mu, \rm G} = \rm const$. For the $\gnl$ case, there is no $\ell_3$ dependence except in the delta function, thus, the integral part is reduced to
\begin{equation}
  \int_2^{\ell_{\rm max}} \ell_1 d\ell_1 \, C_{\ell_1, \rm SW}^{TT} \int_2^{\ell_{\rm max}} \ell_2 d\ell_2 \, C_{\ell_2, \rm SW}^{TT} ~.
\end{equation}
After computing this, we finally obtain
\begin{eqnarray}
  \Delta \gnl|_{\taunl = 0} \simeq
\left( \frac{C_{\ell}^{\mu\mu, \rm G}}{10^{-30}} \right)^{1/2}
  \left[ \ln\left(\frac{\ell_{\rm max}}{2} \right) \right]^{-1} ~. \label{eqn:dgNL_tau0}
\end{eqnarray}
For the $\taunl$ case, the Fisher matrix is proportional to
\begin{eqnarray}
  \left[\prod_{n=1}^3 \int d^2 \boldsymbol{\ell}_n \right] \delta^{(2)}\left( \boldsymbol{\ell}_1 + \boldsymbol{\ell}_2 + \boldsymbol{\ell}_3  \right) 
  \left( \frac{C_{\ell_1, \rm SW}^{TT}}{C_{\ell_2, \rm SW}^{TT}}
+ 2 + \frac{C_{\ell_2, \rm SW}^{TT}}{C_{\ell_1, \rm SW}^{TT}}
  \right) \left( C_{\ell_3, \rm SW}^{TT} \right)^2 ~.
  \end{eqnarray}
The signals satisfying $\ell_3 \ll \ell_1, \ell_2$ contribute dominantly to the integrals and hence we can evaluate this as
\begin{equation}
  16\pi^2 \int_2^{\ell_{\rm max}} \ell_1 d\ell_1 \int_2^{\ell_{\rm max}} \ell_3 d\ell_3 \left( C_{\ell_3, \rm SW}^{TT} \right)^2 ~.
\end{equation}
For large $\ell_{\rm max}$, this is proportional to $\ell_{\rm max}^2$ and we finally have
\begin{eqnarray}
  \Delta \taunl|_{\taunl = 0} \simeq
\left( \frac{C_{\ell}^{\mu\mu, \rm G}}{10^{-30}} \right)^{1/2}
  \frac{5.5}{\ell_{\rm max}} ~.
  \label{eqn:dtauNL_tau0}
\end{eqnarray}
For $\ell_{\rm max} \gtrsim 10$, the analytic expressions \eqref{eqn:dgNL_tau0} and \eqref{eqn:dtauNL_tau0} are in excellent agreement with the numerical results (corresponding to the red dashed lines in Fig.~\ref{fig:dgNL_dtauNL_CV}). On the other hand, for $\taunl \ne 0$, $\Delta \gnl$ deviates drastically from $\propto 1/\ln(\ell_{\rm max} / 2)$, because of non-negligible contributions of $C_\ell^{\mu\mu, \taunl}$ to the denominator of the Fisher matrix.

For comparison, in Fig.~\ref{fig:dgNL_dtauNL_CV}, we also plot our expected uncertainties estimated in a noiseless, cosmic-variance dominated measurement of the CMB temperature trispectrum ($TTTT$), which agree with results in previous literature \cite{Kogo:2006kh,Pearson:2012ba,Shiraishi:2013oqa,Regan:2010cn,Sekiguchi:2013hza}. This level of sensitivity is essentially already achieved using current {\it Planck} data \cite{Ade:2015ava,Feng:2015pva}. As shown in this figure, since the cosmic variance uncertainty for $\mu$-distortions is smaller than that for temperature anisotropies (i.e., $C_\ell^{\mu\mu, \rm G} + C_\ell^{\mu\mu, \taunl} \ll C_\ell^{TT}$), $TT\mu$ allows to achieve better sensitivity to both $\gnl$ and $\taunl$ than $TTTT$ does, for $\ell_{\rm max} \leq 1000$. However, given the difference in scaling with $\ell_{\rm max}$ of the two quantities -- i.e. $\Delta \taunl^{TT\mu} \propto \ell_{\rm max}^{-1}$ \eqref{eqn:dtauNL_tau0} vs. $\Delta \taunl^{TTTT} \propto \ell_{\rm max}^{-2}$ \cite{Kogo:2006kh} -- $TTTT$ might become better than $TT\mu$ at measuring $\taunl$ for higher $\ell_{\rm max}$, and large values of $\taunl$.

\subsection{Effects of experimental uncertainties}

\begin{figure}[t!]
  \begin{tabular}{c}
    \begin{minipage}{1.0\hsize}
  \begin{center}
    \includegraphics[width = 0.8\textwidth]{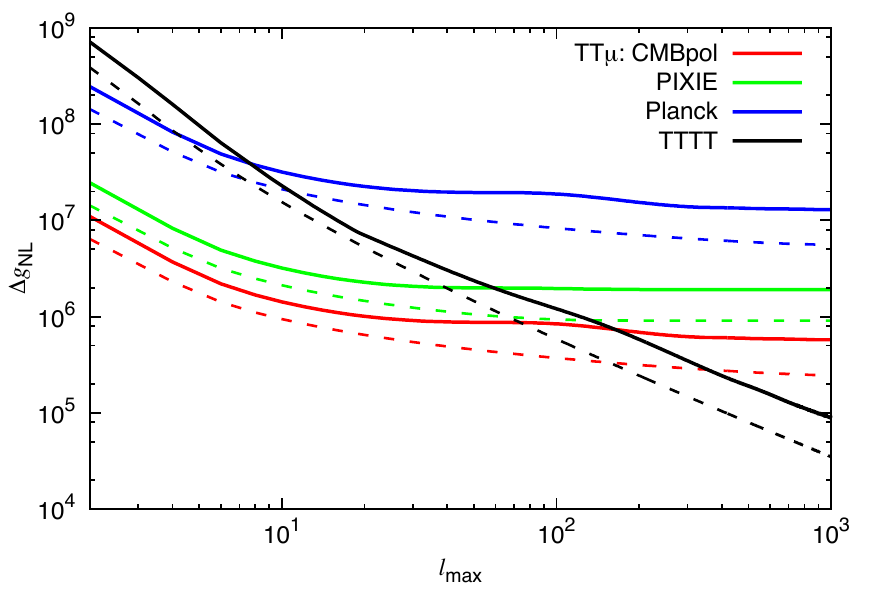}
  \end{center}
\end{minipage}
\end{tabular}
  \caption{Expected $1\sigma$ errors on $\gnl$ computed from $TT\mu$ (colored lines) for noise-levels representative of {\it Planck}, PIXIE and CMBpol. For comparison, we also plot the errors computed from $TTTT$ (black lines) for a noiseless CMB survey, which are almost the same as the errors obtained in the {\it Planck} temperature data analysis \cite{Ade:2015ava,Feng:2015pva}. Solid and dashed lines correspond to the results including full CMB transfer function (Eqs.~\eqref{eqn:redgnl_fisher} and \eqref{eqn:redtau_fisher} for $TT\mu$) and those in the SW limit (Eqs.~\eqref{eqn:TTmu_gNL_SW_approx} and \eqref{eqn:TTmu_tau_SW_approx} for $TT\mu$), respectively. We here assume $\fnl = \taunl = 0$. For $\ell_{\rm max} \lesssim \ell_\mu$, the scalings agree with expectations from Eqs.~\eqref{eqn:Fisher} and \eqref{eqn:dgNL_tau0}: $\Delta \gnl \simeq (N_\mu / 10^{-30})^{1/2} [ \ln(\ell_{\rm max} / 2) ]^{-1}$. At larger $\ell_{\rm max}$, when $N_{\mu}$ starts dominating, the $TT\mu$ sensitivity falls below $TTTT$.} \label{fig:dgNL_tauNL0_noise}
\end{figure}

\begin{figure}
\begin{tabular}{c}
    \begin{minipage}{1.0\hsize}
  \begin{center}
    \includegraphics[width = 0.8\textwidth]{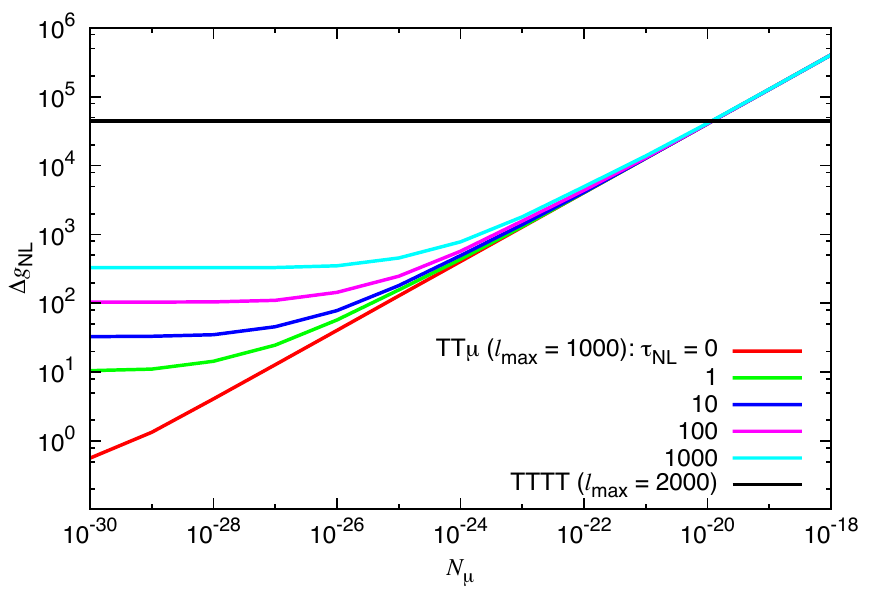}
  \end{center}
\end{minipage}
\end{tabular}
\\
\begin{tabular}{c}
    \begin{minipage}{1.0\hsize}
  \begin{center}
    \includegraphics[width = 0.8\textwidth]{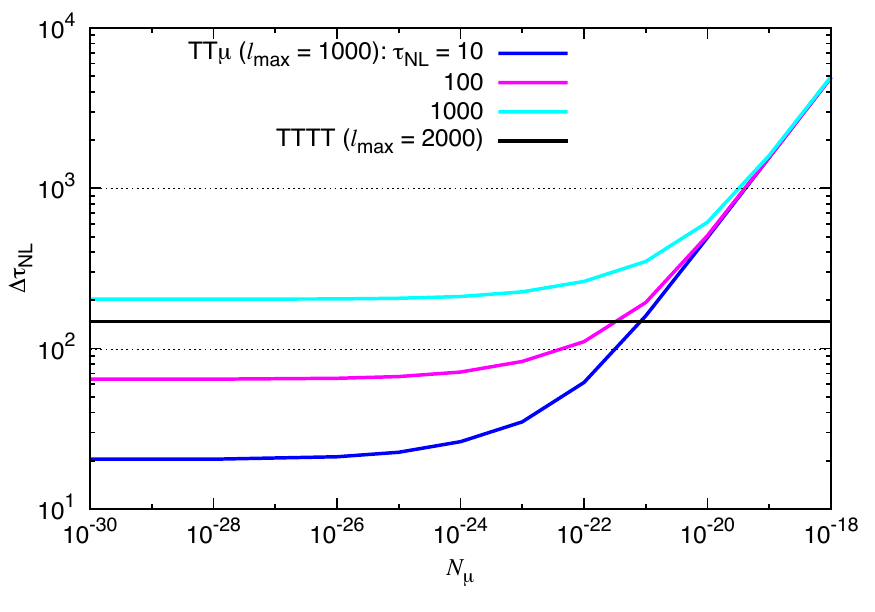}
  \end{center}
\end{minipage}
\end{tabular}
\caption{Expected 1$\sigma$ errors on $\gnl$ (top panel) and $\taunl$ (bottom panel) estimated from $TT\mu$ (colored lines) at $\ell_{\rm max} = 1000$, as a function of the magnitude of instrumental noise $N_\mu$, keeping $\ell_{\mu} = 1000$ angular resolution fixed. Black lines show the expected errors, at $\ell_{\rm max} = 2000$, obtained from $TTTT$ in a noiseless CMB measurement, very close to the error bars obtained from the {\it Planck} temperature data \cite{Ade:2015ava,Feng:2015pva}. In the $TT\mu$ cases, we consider several nonzero $\taunl$'s with $\fnl = 0$. The $TT\mu$ bispectrum used in this estimation is computed from Eqs.~\eqref{eqn:redgnl_fisher} and \eqref{eqn:redtau_fisher}, including the full CMB transfer function dependence.} \label{fig:Nmu_vs_dgNL_dtauNL}
\end{figure}

Besides the ideal, cosmic-variance dominated case, we consider also several different noise levels, corresponding to experiments like {\it Planck} \cite{Planck:2006aa}, PIXIE \cite{Kogut:2011xw} and CMBpol \cite{Baumann:2008aq}. For $\mu$-$\mu$ noise spectra, we assume $N_\ell^{\mu \mu} = N_\mu \exp\left(\ell^2 / \ell_\mu^2 \right)$, with $(N_\mu, \ell_\mu) = (10^{-15}, 861)$ ({\it Planck}), $(10^{-17}, 84)$ (PIXIE) and $(2\times 10^{-18}, 1000)$ (CMBpol) \cite{Ganc:2012ae,Ganc:2014wia}. As it is typical for this type of analysis, we see that current and forthcoming surveys, such as $\Planck$ and PIXIE, are expected to produce error bars on relevant NG parameters which are much worse than what is achievable with the current {\it Planck} measurements or cosmic-variance dominated CMB measurements (compare colored lines with black lines in Fig.~\ref{fig:dgNL_tauNL0_noise}). If we focus on $\gnl$, and consider the fiducial case $\taunl=0$ (resulting in $C_\ell^{\mu\mu} = C_\ell^{\mu\mu, \rm G} + N_\ell^{\mu\mu}$), we find that $\Planck$ can achieve a level of sensitivity $\Delta \gnl \simeq 10^7$, while PIXIE and CMBpol are expected to reach $\Delta \gnl \simeq 2 \times 10^6$ and $\Delta \gnl \simeq 6 \times 10^5$, respectively, as described in Fig.~\ref{fig:dgNL_tauNL0_noise}.

It is interesting to estimate the noise level in $\mu$-distortion measurements, required for $TT\mu$ to achieve better sensitivity than $TTTT$. To this purpose, we compute $\Delta \gnl$ and $\Delta \taunl$ at $\ell_{\rm max} = 1000$, gradually decreasing the magnitude of instrumental noise, $N_\mu$, from $10^{-18}$ to $10^{-30}$. For the angular resolution, we consider $\ell_{\mu} = 1000$, comparable to the value in CMBpol \cite{Baumann:2008aq}. Figure~\ref{fig:Nmu_vs_dgNL_dtauNL} describes our numerical results. For large $N_\mu$, $C_\ell^{\mu\mu}$, at the denominator of \eqref{eqn:Fisher}, is dominated by instrumental noise. The error bars thus scale like $\sqrt{N_\mu}$. However, as $N_\mu$ decreases, $N_\ell^{\mu\mu}$ becomes subdominant compared with $C_\ell^{\mu\mu, \taunl}$ or $C_\ell^{\mu\mu, \rm G}$, and the error bars finally plateau for $N_\mu \lesssim C_\ell^{\mu\mu, \rm G} \sim 10^{-30}$. This behavior is displayed in Fig.~\ref{fig:Nmu_vs_dgNL_dtauNL}, for several fiducial values of $\taunl$. We find from the top panel of Fig.~\ref{fig:Nmu_vs_dgNL_dtauNL} that, if we want $TT\mu$ to outperform $TTTT$ at measuring $\gnl$, $N_\mu \lesssim 10^{-20}$ is required, independently of $\taunl$. In contrast, for the $\taunl$ case, the final result depends strongly on the actual value of $\taunl$. We have already seen in the previous subsection that a detectable $\taunl$ should obey $\Delta \taunl < \taunl$, making $\taunl \sim 40$ the smallest detectable value, for $N_\mu = 0$.
For this reason $TT\mu$ is not useful for measuring small values of $\taunl$. Nonetheless, as seen in the bottom panel of Fig.~\ref{fig:Nmu_vs_dgNL_dtauNL}, $\taunl \sim 100$ is detectable using $TT\mu$, when  $N_\mu \lesssim 10^{-23}$, while being unmeasurable with $TTTT$. If we further increase $\taunl$ to reach $\taunl \sim 1000$, then  $TTTT$ outperforms $TT\mu$. Of course, the most powerful way to measure $\taunl$ using spectral distortions is via $\mu\mu$ correlations \cite{Pajer:2012vz}, and that approach can potentially vastly outperform the temperature trispectrum. If we consider $\taunl$, $TT\mu$ can be essentially used for a cross-check of tighter $\mu \mu$ results.

\section{Conclusions}\label{sec:conclusions}

In this paper, we studied the $TT\mu$ three-point function, arising from correlations between CMB temperature anisotropies and chemical potential ($\mu$) 
distortions in presence of local primordial NG. We showed,  first with a more intuitive argument, followed by a full calculation, that {\em both} $\taunl$ and $\gnl$-type primordial trispectrum signatures source the $TT\mu$ bispectrum. 
Measurements of $TT\mu$ would thus allow to constrain also the $\gnl$ parameter, 
 contrary to what happens with the $\mu\mu$ two-point auto-correlation, which is sensitive only to the $\taunl$ signal \cite{Pajer:2012vz}. 
Our Fisher matrix-based forecast, in line with previous $T\mu$ and $\mu\mu$ analyses \cite{Pajer:2012vz,Ganc:2012ae,Biagetti:2013sr, Miyamoto:2013oua, Kunze:2013uja, Ganc:2014wia, Ota:2014iva, Emami:2015xqa, Shiraishi:2015lma}, shows that an ideal, cosmic-variance dominated experiment 
could in principle determine $\gnl$ and $\taunl$ from $TT\mu$ with an impressive level of accuracy, allowing to detect $\gnl \sim 0.4$ and $\taunl \sim 40$. 
While this is obviously a futuristic scenario, it does reflect the fact that correlations between CMB anisotropies and spectral distortions, including two and three-point functions, contain a large amount of information (since they can probe a vast range of otherwise unaccessible scales), and have the potential to significantly improve on current primordial NG experimental bounds. 

The exact shape of the $TT\mu$ bispectrum should depend on the primordial inflationary scenario under exam. For example, in an inflationary model where a vector field acts as a strong NG source (e.g., \cite{Bartolo:2011ee,Bartolo:2012sd,Bartolo:2015dga}), a direction dependence of the vector field in the curvature trispectrum may give nontrivial effects in $TT\mu$, as well as it does for $TTT$ \cite{Bartolo:2011ee, Shiraishi:2011ph, Shiraishi:2013vja}, $TTTT$ \cite{Shiraishi:2013oqa} and $T\mu$ \cite{Shiraishi:2015lma}. Moreover, $TT\mu$ could also arise from different generation mechanisms. Heating sources, such as magnetic fields stretched on cosmological scales, can generate $T$ and $\mu$ fields which differs from the standard adiabatic mode considered in this paper (e.g., \cite{Kunze:2013uja, Miyamoto:2013oua, Ganc:2014wia}). Also in this case one expects a $TT\mu$ signature with a specific shape.
$TT\mu$ could be interesting also to check alternative models to Inflation. For example, it was argued that ekpyrosis produces a local trispectrum with $\gnl \leq -1700$ \cite{Lehners:2013cka} or $ -1000 \leq \gnl \leq - 100$ \cite{Fertig:2015ola}, a value which is well below the sensitivity of CMB temperature trispectrum measurements, but that is in principle accessible with $TT\mu$ in the future, according to our results. As a starting point, this paper analyzed the most standard case, namely, $TT\mu$ sourced by the adiabatic mode due to the standard $\gnl$ and $\taunl$-type trispectra. Other possibilities, mentioned above, are interesting to investigate and will be accounted for in future works. 
\\



\acknowledgments
 We thank Marc Kamionkowski, Eiichiro Komatsu, Antony Lewis and Sabino Matarrese for useful discussions. MS was supported in part by a Grant-in-Aid for JSPS Research under Grants No.~27-10917, and in part by the World Premier International Research Center Initiative (WPI Initiative), MEXT, Japan. This work was supported in part by ASI/INAF Agreement I/072/09/0 for the Planck LFI Activity of Phase E2.



\bibliography{paper}
\end{document}

%% file: macros.tex
\newcommand{\fnl}{f_{\rm{NL}}}
\newcommand{\fnlloc}{f^{\rm local}_{\rm NL}}
\newcommand{\fnleqi}{f^{\rm equil}_{\rm NL}}
\newcommand{\taunl}{\tau_{\rm{NL}}}
\newcommand{\gnl}{g_{\rm{NL}}}
\newcommand{\htaunl}{\hat{\tau}_{\rm NL}}
\newcommand{\barmodulation}{\bar{f}}
\newcommand{\hatmodulation}{\hat{f}}
\newcommand{\mfmodulation}{\barmodulation^{\rm MF}}
\newcommand{\hatReconNoise}{\hat{N}_L}
\providecommand{\fsky}{f_{\rm sky}}
\newcommand{\fskyeff}{\fsky^{\rm eff}}

\providecommand{\planck}{\textit{Planck}}

\newcommand{\covinvT}{\bar{T}}
\newcommand{\Ltemp}{\tilde{T}}
\newcommand{\lensedC}{\tilde{C}}
\newcommand{\Lmin}{{L_{\rm min}}}
\newcommand{\Lmax}{{L_{\rm max}}}

\providecommand{\alt}{\lea}
\providecommand{\agt}{\gea}

\providecommand{\LCDM}{{$\rm{\Lambda CDM}$}}

\newcommand\ba{\begin{eqnarray}}
\newcommand\ea{\end{eqnarray}}
\newcommand\bea{\begin{eqnarray}}
\newcommand\eea{\end{eqnarray}}

\newcommand\be{\begin{equation}}
\newcommand\ee{\end{equation}}

\providecommand{\var}{\text{var}}
\providecommand{\cov}{\text{cov}}

\newcommand{\ud}{{\rm d}}



\newcommand{\boldvec}[1]{{{\vec{#1}}}}

\newcommand{\vA}{\boldvec{A}}
\newcommand{\vB}{\boldvec{B}}
\newcommand{\vC}{\boldvec{C}}
\newcommand{\vD}{\boldvec{D}}
\newcommand{\vE}{\boldvec{E}}
\newcommand{\vF}{\boldvec{F}}
\newcommand{\vG}{\boldvec{G}}
\newcommand{\vH}{\boldvec{H}}
\newcommand{\vI}{\boldvec{I}}
\newcommand{\vJ}{\boldvec{J}}
\newcommand{\vK}{\boldvec{K}}
\newcommand{\vL}{\boldvec{L}}
\newcommand{\vM}{\boldvec{M}}
\newcommand{\vN}{\boldvec{N}}
\newcommand{\vO}{\boldvec{O}}
\newcommand{\vP}{\boldvec{P}}
\newcommand{\vQ}{\boldvec{Q}}
\newcommand{\vR}{\boldvec{R}}
\newcommand{\vS}{\boldvec{S}}
\newcommand{\vT}{\boldvec{T}}
\newcommand{\vU}{\boldvec{U}}
\newcommand{\vV}{\boldvec{V}}
\newcommand{\vW}{\boldvec{W}}
\newcommand{\vX}{\boldvec{X}}
\newcommand{\vY}{\boldvec{Y}}
\newcommand{\vZ}{\boldvec{Z}}

\newcommand{\va}{\boldvec{a}}
\newcommand{\vb}{\boldvec{b}}
\newcommand{\vc}{\boldvec{c}}
\newcommand{\vd}{\boldvec{d}}
\newcommand{\ve}{\boldvec{e}}
\newcommand{\vf}{\boldvec{f}}
\newcommand{\vg}{\boldvec{g}}
\newcommand{\vh}{\boldvec{h}}
\newcommand{\vi}{\boldvec{i}}
\newcommand{\vj}{\boldvec{j}}
\newcommand{\vk}{\boldvec{k}}
\newcommand{\vl}{\boldvec{l}}
\newcommand{\vm}{\boldvec{m}}
\newcommand{\vn}{\boldvec{n}}
\newcommand{\vo}{\boldvec{o}}
\newcommand{\vp}{\boldvec{p}}
\newcommand{\vq}{\boldvec{q}}
\newcommand{\vs}{\boldvec{s}}
\newcommand{\vt}{\boldvec{t}}
\newcommand{\vu}{\boldvec{u}}
\newcommand{\vv}{\boldvec{v}}
\newcommand{\vw}{\boldvec{w}}
\newcommand{\vx}{\boldvec{x}}
\newcommand{\vy}{\boldvec{y}}
\newcommand{\vz}{\boldvec{z}}

\newcommand{\cla}{\mathcal{A}}
\newcommand{\clb}{\mathcal{B}}
\newcommand{\clc}{\mathcal{C}}
\newcommand{\cld}{\mathcal{D}}
\newcommand{\cle}{\mathcal{E}}
\newcommand{\clf}{\mathcal{F}}
\newcommand{\clg}{\mathcal{G}}
\newcommand{\clh}{\mathcal{H}}
\newcommand{\cli}{\mathcal{I}}
\newcommand{\clj}{\mathcal{J}}
\newcommand{\clk}{\mathcal{K}}
\newcommand{\cll}{\mathcal{L}}
\newcommand{\clm}{\mathcal{M}}
\newcommand{\cln}{\mathcal{N}}
\newcommand{\clo}{\mathcal{O}}
\newcommand{\clp}{\mathcal{P}}
\newcommand{\clq}{\mathcal{Q}}
\newcommand{\clr}{\mathcal{R}}
\newcommand{\cls}{\mathcal{S}}
\newcommand{\clt}{\mathcal{T}}
\newcommand{\clu}{\mathcal{U}}
\newcommand{\clv}{\mathcal{V}}
\newcommand{\clw}{\mathcal{W}}
\newcommand{\clx}{\mathcal{X}}
\newcommand{\cly}{\mathcal{Y}}
\newcommand{\clz}{\mathcal{Z}}

\newcommand{\vnhat}{\hat{\vn}}
\newcommand{\vrhat}{\hat{\vr}}
\newcommand{\vkhat}{\hat{\vk}}

\def\Commander{\texttt{Commander}}
\def\Ruler{\texttt{Ruler}}
\def\CR{\texttt{C-R}}
\def\NILC{\texttt{NILC}}
\def\SMICA{\texttt{SMICA}}
\def\SEVEM{\texttt{SEVEM}}

\def\hf{\frac{1}{2}}
\def\pl{\parallel}
\def\pp{\bot}
\def\px{\approx}
\def\pr{\prime}

\def\={\nonumber &=}
\def\nn{\nonumber}

\def\({\left(}
\def\){\right)}
\def\[{\left[}
\def\]{\right]}
\def\<{\left\langle}
\def\>{\right\rangle}

\def\uk{{\bf \hat{k}}}
\def\un{{\bf \hat{n}}}
\def\ur{{\bf \hat{r}}}
\def\ux{{\bf \hat{x}}}
\def\bk{{\bf k}}
\def\bn{{\bf n}}
\def\br{{\bf r}}
\def\bv{{\bf v}}
\def\bx{{\bf x}}
\def\bl{{\bf l}}
\def\bkp{{\bf k^\pr}}
\def\brp{{\bf r^\pr}}

\def\bib{\bibitem}
\def\curl{\mathcal}
\def\nl{\nonumber &=&}
\def\eq{\begin{eqnarray}}
\def\qe{\end{eqnarray}}

\def\and{\quad \mbox{and} \quad}
\def\imp{\quad \Rightarrow \quad}

\newenvironment{mylist}{\begin{list}
    {$\bullet$}
    {\setlength{\leftmargin}{1cm}
    \setlength{\rightmargin}{1cm}
    \setlength{\labelsep}{.5cm}
    \setlength{\itemsep}{.5cm}}}
{\end{list}}

\def\Planck{\textit{Planck}}

\def\fnl{f_\textrm{NL}}
\def\gnl{g_\textrm{NL}}
\def\taunl{\tau_\textrm{NL}}
\def\barfnl{\bar f_\textrm{NL}}
\def\Fnl{ F_\textrm {NL}}
\def\barFnl{ \bar F_\textrm {NL}}
\def\Fnllocal{F_\textrm {NL}^\textrm{local}}
\def\fnllocal{f_\textrm {NL}^\textrm{local}}
\def\Fnlequil{F_\textrm {NL}^\textrm{equil}}
\def\fnlequil{f_\textrm {NL}^\textrm{equil}}
\def\Fnlequil{F_\textrm {NL}^\textrm{dbi}}
\def\fnlequil{f_\textrm {NL}^\textrm{dbi}}
\def\Fnlfeat{F_\textrm {NL}^\textrm{feat}}
\def\fnlfeat{f_\textrm {NL}^\textrm{feat}}
\def\Fnlmodel{F_\textrm {NL}^\textrm{model}}
\def\fnlmodel{f_\textrm {NL}^\textrm{model}}
\def\Fnlth{F_\textrm {NL}^\textrm{th}}
\def\fnlth{f_\textrm {NL}^\textrm{th}}
\def\bfnl{\kern2pt\overline{\kern-2ptf}_\textrm{NL}}

\def\lmax{\ell_\textrm{max}}
\def\lall{\ell_1,\ell_2,\ell_3}
\def\lalltwo{\ell_4,\ell_5,\ell_6}
\def\lsum{\ell_1+\ell_2+\ell_3}
\def\ltilde{\tilde l}
\def\Blll{B_{\ell_1\ell_2\ell_3}}
\def\blll{b_{\ell_1\ell_2\ell_3}}
\def\bllllocal{b_{\ell_1\ell_2\ell_3}^\textrm{local}}
\def\blllmodel{b_{\ell_1\ell_2\ell_3}^\textrm{model}}
\def\blllconst{b_{\ell_1\ell_2\ell_3}^\textrm{const}}
\def\blllequil{b_{\ell_1\ell_2\ell_3}^\textrm{equil}}
\def\blllstring{b_{\ell_1\ell_2\ell_3}^\textrm{string}}
\def\hlll{h_{\ell_1\ell_2\ell_3}}
\def\hllltwo{h_{\ell_4\ell_5\ell_6}}

\def\th{\textrm{th}}
\def\exp{\textrm{exp}}

\def\kmax{k_\textrm{max}}
\def\kall{k_1,k_2,k_3}
\def\ksum{k_1+k_2+k_3}
\def\ktilde{\tilde k}
\def\bfkall{\textbf{k}_1,\,\textbf{k}_2,\,\textbf{k}_3}
\def\Bkkk{B(\kall)}
\def\Skkk{S(\kall)}

\def\Cl{C_\ell}
\def\alm{a_{\ell m}}
\def\almone{a_{\ell_1m_1}}
\def\almtwo{a_{\ell_2m_2}}
\def\almthree{a_{\ell_3m_3}}
\def\almfour{a_{\ell_4m_4}}
\def\almfive{a_{\ell_5m_5}}
\def\almsix{a_{\ell_6m_6}}
\def\Ylm{Y_{\ell m}}

\def\Vtetra{{{\cal V}_{\cal T}}}

\def\Q{\curl{Q}}
\def\Qn{\curl{Q}_n}
\def\Qm{\curl{Q}_m}
\def\Qp{\curl{Q}_p}
\def\Qnxyz{\curl{Q}_n(x,y,z)}

\def\R{\curl{R}}
\def\Rn{\curl{R}_n}
\def\Rm{\curl{R}_m}
\def\Rp{\curl{R}_p}
\def\Rnxyz{\curl{Q}_n(x,y,z)}

\def\barQ{\kern2pt\overline{\kern-2pt\curl{Q}}}
\def\barQn{\barQ_n}
\def\barQm{\barQ_m}
\def\barQp{\barQ_p}
\def\barQnxyz{\barQ_n(x,y,z)}

\def\bargamma{\kern2pt\overline{\kern-2pt\gamma}}
\def\barzeta{\kern2pt\overline{\kern-2pt\zeta}}

\def\barR{\kern2pt\overline{\kern-2pt\curl{R}}}
\def\barRn{\barR_n}
\def\barRm{\barR_m}
\def\barRp{\barR_p}
\def\barRnxyz{\barR_n(x,y,z)}

\def\nmax{n_\textrm{max}}

\def\aR{\alpha^{\scriptscriptstyle{\cal R}}}
\def\aRn{\aR_n}
\def\aRp{\aR_p}
\def\baR{\bar{\alpha}^{\scriptscriptstyle{\cal R}}}
\def\baRn{\baR_n}
\def\baRp{\baR_p}

\def\aQ{\alpha^{\scriptscriptstyle{\cal Q}}}
\def\aQn{\aQ_n}
\def\aQp{\aQ_p}
\def\baQ{\bar{\alpha}^{\scriptscriptstyle{\cal Q}}}
\def\baQn{\baQ_n}
\def\baQp{\baQ_p}

\def\bR{\beta^{\scriptscriptstyle{\cal R}}}
\def\bRn{\bR_n}
\def\bRp{\bR_p}
\def\bbR{\bar{\beta}^{\scriptscriptstyle{\cal R}}}
\def\bbRn{\bbR_n}
\def\bbRp{\bbR_p}

\def\bQ{\beta^{\scriptscriptstyle{\cal Q}}}
\def\bQn{\bQ_n}
\def\bQp{\bQ_p}
\def\bbQ{\bar{\beta}^{\scriptscriptstyle{\cal Q}}}
\def\bbQn{\bbQ_n}
\def\bbQp{\bbQ_p}

\def\Qndefn{q_{\{p}\,q_r\,q_{s\}}}
\def\bQndefn{\bar q_{\{p}\,\bar q_r\,\bar q_{s\}}}

\def\balpha{\mbox{\boldmath$\alpha$}_l}
\def\bbeta{\mbox{\boldmath$\beta$}_l}
\def\bgamma{\mbox{\boldmath$\gamma$}_l}
\def\bdelta{\mbox{\boldmath$\delta$}_l}
\def\Malpha{M_\alpha(x,\hat{\bf n})}
\def\Mbeta{M_\beta(x,\hat{\bf n})}
\def\Mgamma{M_\gamma(x,\hat{\bf n})}
\def\Mdelta{M_\delta(x,\hat{\bf n})}

\newcommand{\kv}{{\bf k}}
\def\alll{\ell_1 \ell_2 \ell_3}
\def\fnll{f_{\rm NL}^{\rm local}}

\def\threej#1#2#3#4#5#6{\left( \begin{array}{ccc} #1 & #2 & #3 \\ #4 & #5 & #6 \end{array} \right) }
\def\llist{\ell_1,\ell_2,\ell_3}
\def\klist{k_1,k_2,k_3}
\def\eqref#1{(\ref{#1})}

\def\leaderfi1{\leaders\hbox to 5pt{\hss.\hss}\hfil}


\def\setsize{\csname @setfontsize\endcsname \setsize}